\documentstyle[12pt,epsfig]{article}  
\begin{document} 
\baselineskip=18pt

\def\la{\mathrel{\mathpalette\fun <}}
\def\ga{\mathrel{\mathpalette\fun >}}
\def\fun#1#2{\lower3.6pt\vbox{\baselineskip0pt\lineskip.9pt
\ialign{$\mathsurround=0pt#1\hfil##\hfil$\crcr#2\crcr\sim\crcr}}} 

\begin{titlepage} 
\begin{flushright}{\bf Preprint SINP MSU 2004-13/752}\end{flushright}
\begin{center}
$~$ \vskip 5 cm 
{\Large \bf Fast simulation of jet quenching in ultrarelativistic heavy ion 
collisions} \\

\vspace{10mm}

I.P.~Lokhtin and A.M.~Snigirev  \\ ~ \\ 
M.V.Lomonosov Moscow State University, D.V.Skobeltsyn Institute of Nuclear Physics \\
119992, Vorobievy Gory, Moscow, Russia \\ E-mail:~~Igor.Lokhtin@cern.ch   
\end{center}  

\vspace{5mm} 

\begin{abstract} 
The method for simulation of medium-induced rescattering and energy loss of hard
partons in ultrarelativistic heavy ion collisions is developed. The model is 
realized as fast Monte-Carlo tool implemented to modify standard PYTHIA jet 
event.  
\end{abstract}

\end{titlepage}   

\section{Introduction} 

The experimental investigation of ultra-relativistic nuclear collisions offers 
a unique possibility of studying the properties of strongly interacting matter 
at high energy density. In that regime, hadronic matter is expected to become 
deconfined, and a gas of asymptotically free quarks and gluons is formed, the 
so-called quark-gluon plasma (QGP), in which the colour interactions between 
partons are screened owing to collective effects~\cite{qm}. One of the important 
tools to study QGP properties in heavy ion collisions is a QCD jet production. 
Medium-induced energy loss of energetic partons, the so-called jet quenching, 
has been proposed to be very different in cold nuclear matter and in QGP, 
resulting in many challenging observable phenomena~\cite{baier_rev}. Recent RHIC 
data on suppression of inclusive high-p$_T$ charge and neutral hadron production 
from STAR~\cite{star}, PHENIX~\cite{phenix}, PHOBOS~\cite{phobos} and 
BRAHMS~\cite{brahms} are in agreement with the jet quenching 
hypothesis~\cite{Wang:2004}. However direct event-by-event reconstruction of 
jets and their characteristics is not available in RHIC experiments at the 
moment, while the assumption that integrated yield of all high-$p_T$ particles 
originates only from jet fragmentation is not obvious.

At LHC a new regime of heavy ion physics will be reached at 
$\sqrt{s_{\rm NN}}=5.5$ TeV where hard and semi-hard QCD multi-particle 
production can dominate over underlying soft events. The initial gluon densities 
in Pb$-$Pb reactions at LHC are expected to be significantly higher than at 
RHIC, implying stronger partonic energy loss which can be observable in various 
new channels~\cite{lhc-jets}. Thus in order to test a 
sensitivity of accessible at LHC observables to jet quenching, and to study
corresponding experimental capabilities of real detectors, the development of 
fast Monte-Carlo tools is necessary. 

\section{Physics frameworks of the model} 

The detailed description of physics frameworks of the developed model can be
found in a number of our previous 
papers~\cite{lokhtin98,lokhtin00,lokhtin01,lokhtin02,lokhtin03}. 
The approach relies on an accumulative energy losses, when gluon radiation is 
associated with each scattering in expanding medium together including the 
interference effect by the modified radiation spectrum $dE/dl$ as a function 
of decreasing temperature $T$. The basic kinetic integral equation for the 
energy loss $\Delta E$ as a function of initial energy $E$ and path length $L$ 
has the form 
\begin{eqnarray} 
\label{elos_kin}
\Delta E (L,E) = \int\limits_0^Ldl\frac{dP(l)}{dl}
\lambda(l)\frac{dE(l,E)}{dl} \, , ~~~~ 
\frac{dP(l)}{dl} = \frac{1}{\lambda(l)}\exp{\left( -l/\lambda(l)\right) }
\, ,  
\end{eqnarray} 
where $l$ is the current transverse coordinate of a parton, $dP/dl$ is the 
scattering probability density, $dE/dl$ is the energy loss per unit length, 
$\lambda = 1/(\sigma \rho)$ is in-medium mean free path, $\rho \propto T^3$ is 
the medium density at the temperature $T$, $\sigma$ is the integral cross 
section of parton interaction in the medium. 

The collisional energy loss due to elastic scattering with high-momentum 
transfer have been originally estimated by Bjorken in~\cite{bjork82}, and 
recalculated later in~\cite{mrow91} taking also into account the loss with 
low-momentum transfer dominated by the interactions with plasma collective 
modes. Since latter process contributes to the total collisional loss 
without the large factor $\sim \ln{(E / \mu_D)}$ ($\mu_D$ is the Debye 
screening mass) in comparison with high-momentum scattering and it can be 
effectively ``absorbed'' by the redefinition of minimal momentum transfer 
$t \sim \mu_D^2$ under the numerical estimates, we used the collisional 
part with high-momentum transfer only~\cite{lokhtin00},   
\begin{equation} 
\label{col} 
\frac{dE}{dl}^{col} = \frac{1}{4T \lambda \sigma} 
\int\limits_{\displaystyle \mu^2_D}^
{\displaystyle t_{\rm max}}dt\frac{d\sigma }{dt}t ~,
\end{equation} 
and the dominant contribution to the differential cross section 
\begin{equation} 
\label{sigt} 
\frac{d\sigma }{dt} \cong C \frac{2\pi\alpha_s^2(t)}{t^2} 
\frac{E^2}{E^2-m_q^2}~,~~~~
\alpha_s = \frac{12\pi}{(33-2N_f)\ln{(t/\Lambda_{QCD}^2)}} \>
\end{equation} 
for scattering of a parton with energy $E$ off the ``thermal'' partons with 
energy (or effective mass) $m_0 \sim 3T \ll E$. Here $C = 9/4, 1, 4/9$ for $gg$, $gq$ and 
$qq$ scatterings respectively, $\alpha_s$ is the QCD running coupling constant 
for $N_f$ active quark flavors, and $\Lambda_{QCD}$ is the QCD scale parameter 
which is of the order of the critical temperature,  $\Lambda_{QCD}\simeq T_c 
\simeq 200$ MeV. The integrated cross section $\sigma$ is regularized by the 
Debye screening mass squared $\mu_D^2 (T) \simeq 4\pi \alpha _s T^2(1+N_f/6)$. 
The maximum momentum transfer $t_{\rm max}=[ s-(m_p+m_0)^2] [ s-(m_p-m_0)^2 ] / 
s$ where $s=2m_0E+m_0^2+m_p^2$, $m_p$ is the hard parton mass.

There are several calculations of the inclusive energy distribution of
medium-induced gluon radiation from Feyman multiple scattering diagrams. The relation
between these approaches and their main parameters have been discussed in
details in the recent writeup of the working group ``Jet Physics'' for the 
CERN Yellow Report~\cite{lhc-jets}. We restrict to ourself here by using BDMS
formalism~\cite{baier}. In the BDMS framework the strength of multiple
scattering is characterized by the transport coefficient 
$\hat{q}=\mu_D^2/\lambda_g $ ($\lambda_g$ is the gluon mean free path), which is
related to the elastic scattering cross section $\sigma$ (\ref{sigt}). In our
simulations this strength in fact is regulated mainly by the initial QGP 
temperature $T_0$. Then the energy spectrum of coherent medium-induced gluon 
radiation and the corresponding dominated part of radiative energy loss of
massless parton has the form~\cite{baier}: 
\begin{eqnarray} 
\label{radiat} 
\frac{dE}{dl}^{rad} = \frac{2 \alpha_s (\mu_D^2) C_R}{\pi L}
\int\limits_{\omega_{\min}}^E  
d \omega \left[ 1 - y + \frac{y^2}{2} \right] 
\>\ln{\left| \cos{(\omega_1\tau_1)} \right|} 
\>, \\  
\omega_1 = \sqrt{i \left( 1 - y + \frac{C_R}{3}y^2 \right)   
\bar{\kappa}\ln{\frac{16}{\bar{\kappa}}}}
\quad \mbox{with}\quad 
\bar{\kappa} = \frac{\mu_D^2\lambda_g  }{\omega(1-y)} ~, 
\end{eqnarray} 
where $\tau_1=L/(2\lambda_g)$, $y=\omega/E$ is the fraction of the hard parton energy 
carried by the radiated gluon, and $C_R = 4/3$ is the quark color factor. 
A similar expression for the gluon jet can be obtained by substituting 
$C_R=3$ and a proper change of the factor in the square bracket in (\ref{radiat}), see
ref.~\cite{baier}. The integral (\ref{radiat}) is carried out over all energies from 
$\omega_{\min}=E_{LPM}=\mu_D^2\lambda_g$, the minimal radiated gluon energy in 
the coherent LPM regime, up to initial jet energy $E$. 

The generalization of the formula for heavy quark of mass $m_q$ was done by 
using ``dead-cone'' approximation~\cite{dc}:  
\begin{equation}
\label{radmass} 
\frac{dE}{dx}| _{m_q \ne 0} =  \frac{1}{(1+(l\omega )^{3/2})^2}
\frac{dE}{dx}| _{m_q=0}, ~~~ l=\left( \frac{\lambda}{\mu_D^2}\right) ^{1/3}
\left( \frac{m_q}{E}\right) ^{4/3}~, 
\end{equation}
but note there are exist more recent developments on heavy quark energy loss in 
the literature~\cite{djor,armesto}. 

The medium was treated as a boost-invariant longitudinally expanding quark-gluon 
fluid, and partons as being produced on a hyper-surface of equal proper times 
$\tau$~\cite{bjorken}. In order to simplify numerical calculations in original
version of the model we omit the transverse expansion and viscosity of the 
fluid using the well-known scaling Bjorken's solution~\cite{bjorken} for 
temperature and density of QGP at $T > T_c \simeq 200$ MeV:
\begin{equation}
\varepsilon(\tau) \tau^{4/3} = \varepsilon_0 \tau_0^{4/3},~~~~
T(\tau) \tau^{1/3} = T_0 \tau_0^{1/3},~~~~ \rho(\tau) \tau = \rho_0 \tau_0 .
\end{equation}
For certainty we used the initial conditions for the gluon-dominated plasma 
formation expected for central Pb$-$Pb collisions at LHC~\cite{esk}: 
$$\tau_0 \simeq 0.1~{\rm fm/c}, ~~~~T_0 \simeq 1~{\rm GeV}, ~~~~\rho_g 
\approx 1.95T^3 ~.$$ 
Then for non-central collisions and for other beam atomic numbers we suggest 
the proportionality of the initial energy density $\varepsilon _0$ to the ratio 
of nuclear overlap function and effective transverse area of nuclear 
overlapping~\cite{lokhtin00}.

Note, however, that using other initial parameters and scenarious of QGP 
space-time evolution for Monte-Carlo realization of the model is possible 
(by changing some internal parameters of the routine). In fact, the influence of 
the transverse flow, as well as of the mixed phase at $T = T_c$, on the 
intensity of jet rescattering (which is a strongly increasing function of $T$) 
has been found to be inessential for high initial temperatures $T_0 \gg T_c$. 
On the contrary, the presence of QGP viscosity slows down the cooling rate, 
which leads to a jet parton spending more time in the hottest regions of the 
medium. As a result the rescattering intensity goes up, i.e., in fact an 
effective temperature of the medium gets lifted as compared with the perfect QGP 
case. We also do not take into account here the probability of jet rescattering 
in nuclear matter, because the intensity of this process and corresponding 
contribution to total energy loss are not significant due to much smaller 
energy density in a ``cold'' nuclei.

Another important set of the model is the angular spectrum of in-medium gluon 
radiation. Since the full treatment of angular spectrum of emitted gluons is 
rather sophisticated and 
model-dependent~\cite{lokhtin98,baier,Zakharov:1999,urs,vitev}, simple 
parameterizations of gluon angular distribution over emission angle $\theta$ was
used:
\begin{equation} 
\label{sar} 
\frac{dN^g}{d\theta}\propto \sin{\theta} \exp{\left( -\frac{(\theta-\theta
_0)^2}{2\theta_0^2}\right) }~, 
\end{equation}
where $\theta_0 \sim 5^0$ is the typical angle of coherent gluon radiation 
estimated in work~\cite{lokhtin98}. Other parameterizations are also possible. 

\section{Monte-Carlo simulation procedure} 

The model has been realized as fast Monte-Carlo event generator, and
corresponding Fortran routine PYQUEN.F is available by the web~\cite{pyquen}. 
The following input parameters should be specified by user to fix QGP properties: 
beam and target nucleus atomic number and type of event centrality selection  
(options ``fixed impact parameter'' or ``minimum bias events'' are foreseen).   
Since the routine is implemented as a modification of standard PYTHIA6.2 jet
event~\cite{pythia}, the main user program should be compiled with this version of
PYTHIA.  

The following event-by-event Monte-Carlo simulation procedure is applied. 

\begin{itemize} 
\item Generation of initial parton spectra with PYTHIA (fragmentation {\em off}). 
\item Generation of jet production vertex at impact parameter 
$b$ according to the distribution
$$\frac{dN^{\rm jet}}{d\psi dr} (b) = \frac{T_A(r_1) T_A(r_2)}
{\int\limits_0^{2\pi} d \psi \int\limits_0^{r_{max}}r dr T_A(r_1) T_A(r_2)} ,$$
where $r_{1,2} (b,r,\psi)$ are the distances between the nucleus centers and the 
jet production vertex $V(r\cos{\psi}, r\sin{\psi})$; $r_{max} (b, \psi) \le 
R_A$ is the maximum possible transverse distance $r$ from the nuclear collision 
axis to the $V$; $R_A$ is the radius of the nucleus $A$; $T_A(r_{1,2})$ is the 
nuclear thickness function (see ref.~\cite{lokhtin00} for detailed nuclear 
geometry explanations).
\item Generation of scattering cross section $d\sigma / dt$ (\ref{sigt}). 
\item Generation of transverse distance between scatterings, 
$l_i = (\tau_{i+1} - \tau_i)E/p_T$:  
$$ \frac{dP}{dl_i} = \lambda^{-1}(\tau_{i+1}) \exp{(-\int\limits_0^{l_i}
\lambda^{-1} (\tau_i + s)ds)}, ~~~, \lambda^{-1}(\tau ) =\sigma (\tau ) \rho 
(\tau ) .$$ 
\item Reducing parton energy by collisional (\ref{col}) and radiative 
(\ref{radiat}), (\ref{radmass}) loss per scattering $i$:
$$\Delta E_{{\rm tot},i} = \Delta E_{{\rm col},i} + \Delta E_{{\rm rad},i} .$$
\item Calculation of parton transverse momentum kick due to elastic scattering 
$i$:
$$ \Delta k_{t,i}^2 =(E-\frac{t_i}{2m_{0i}})^2-(p-\frac{E}{p}\frac{t_i}{2m_{0i}}-
\frac{t_i}{2p})^2-m_p^2 .$$ 
\item Halting parton rescattering if {\em 1)} parton escapes from dense zone, or 
{\em 2)} QGP cools down to $T_c=200$ MeV, or {\em 3)} parton loses so much 
energy that its $p_T (\tau)$ drops below $2T (\tau)$. 
\item In the end of each event adding new (in-medium emitted) gluons into 
PYTHIA parton list and rearrangements of partons to update string formation are
performed.  
\item Formation of final hadrons by PYTHIA (fragmentation {\em on}).  
\end{itemize} 

As the example, let us compare hadron $p_T$-spectra obtained with jet quenching
(PYTHIA+ PYQUEN) and without one (PYTHIA only). Fig.1 shows this spectrum for 
$\sqrt{s_{\rm pp}}=5.5$ TeV and QGP
parameters selected for central Pb$-$Pb collisions, the events being triggered by having 
at least one jet with $E_T>100$ GeV in the final state. As it could be expected, 
energy loss of hard partons results in the high-$p_T$ suppression of the 
spectrum, while the intensive gluon emission of soft and semi-hard gluons governs 
the low-$p_T$ enhancement. 

\vskip 1 cm 

\begin{figure}[htbp] 
\begin{center} 
\makebox{\epsfig{file=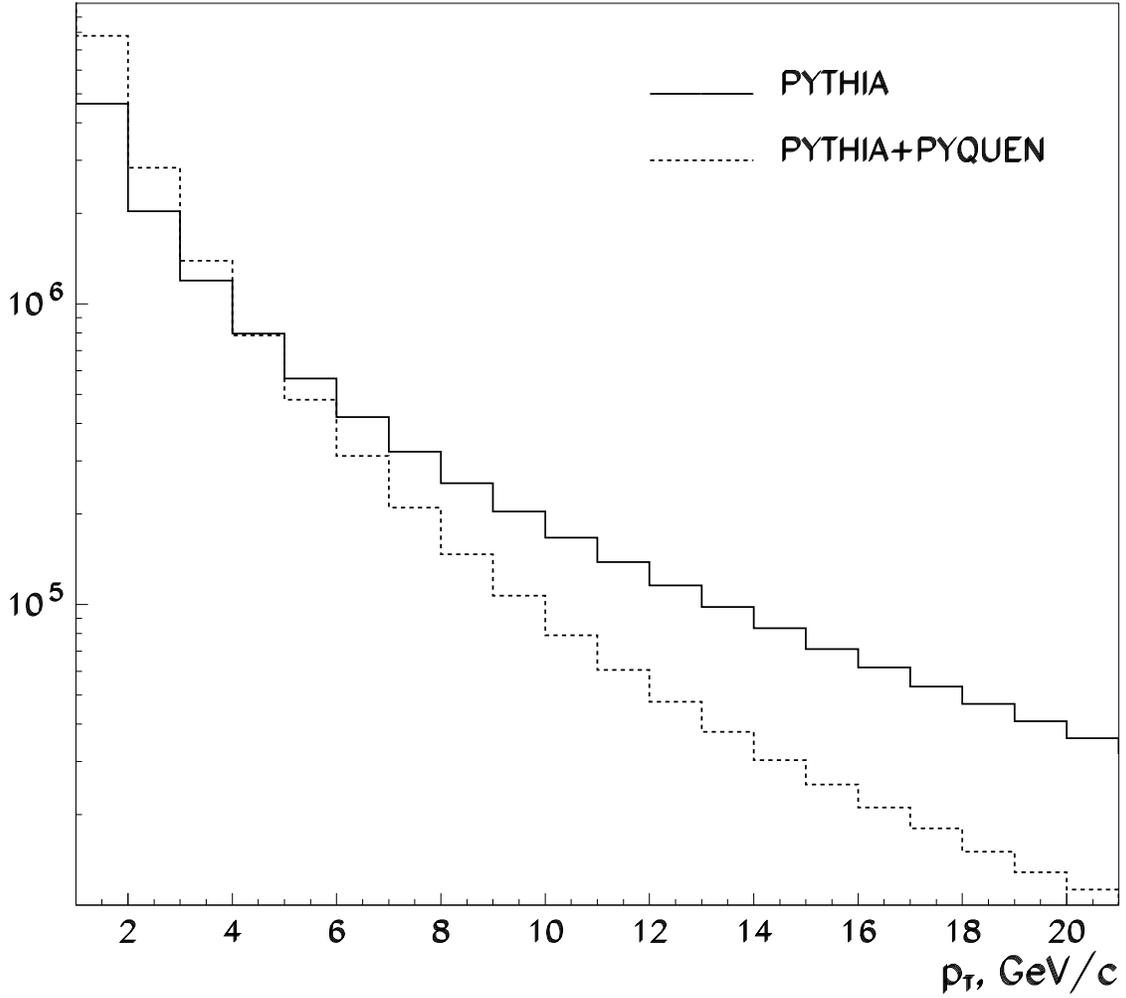, height=160mm}}   
\vskip -5 mm
\caption{\small The hadron $p_T$-spectrum at $\sqrt{s_{\rm pp}}=5.5$ TeV obtained
with jet quenching (PYTHIA+PYQUEN, dashed histogram) and without one (PYTHIA, 
solid histogram). QGP parameters were selected for central Pb$-$Pb collisions. 
Events were triggered having at least one jet with $E_T>100$ GeV in the final 
state.}
\end{center}
\end{figure}

\newpage 
\section{Conclusions} 

The method to simulate rescattering and energy loss of hard partons in 
ultrarelativistic heavy ion collisions has been developed. The model 
is realized as fast Monte-Carlo tool implemented to modify standard PYTHIA jet 
event. Corresponding Fortran routine is available by the web. 
     
To conclude, let us discuss the physics validity of the model application. 
\begin{itemize} 
\item Internal parameters of the routine for initial conditions and space-time
evolution of quark-gluon plasma were selected as an estimation for LHC heavy ion 
beam energies. The result for other beam energy ranges, obtained without 
additional internal parameters adjusting, is not expected to be reasonable. 
\item Hydro-type description of expanding quark-gluon plasma used by the model 
can be applicable for central and semi-central collisions. The result obtained 
for very peripheral collisions ($b \sim 2 R_A$) can be not adequate. 
\item Physics model for medium-induced gluon radiation is valid for relatively 
high transverse momenta of jet partons ($\gg 1$ GeV/$c$). Thus setting 
reasonably high value of minimum $p_T$ in initial hard parton sub-processes 
in PYTHIA is preferable. 
\end{itemize} 

{\it Acknowledgments.} \\
We would like to thank R.~Baier, D.~Denergi, Yu.L.~Dokshitzer, M.~Gyulassy,
A.B.~Kaidalov, N.A.~Kruglov, A.~Morsch, C.~Salgado, L.I.~Sarycheva, D.~Schiff, 
J.~Schukraft, T.~Sjostrand, C.Yu.~Teplov, V.V.~Uzhinskii, I.N.~Vardanian, 
I.~Vitev, R.~Vogt, U.~Wiedemann, B.~Wyslouch, B.G.~Zakharov and G.M.~Zinovjev 
for useful discussions.

\end{document}